\documentstyle[aps,preprint,epsf]{revtex}
\begin{document}
\draft
\title{CONFINEMENT AND SCALING IN DEEP INELASTIC PROCESSES
\thanks{WIS-97/2/Jan.-PH. 
Invited paper given at ISHEPP, Dubna, September 1996.}}
\author{S.A. Gurvitz}

\address{Department of Particle Physics, Weizmann Institute of
         Science, Rehovot 76100, Israel}
\maketitle
\vskip 2cm
The existing data for hadron structure functions, $F_2(x,Q^2)$, 
show considerable $Q^2$-dependence, which is mainly attributed 
to the QCD logarithmic corrections to Bjorken scaling. However, 
at $x\to 1$ the scaling violations are dominated by power corrections 
$\propto 1/Q^2$ (higher twist and target mass effects):
\begin{equation}
F_2(x,Q^2)= F_2^{as}(x,Q^2)+\frac{B(x)}{Q^2}+\cdots,
\label{a1}
\end{equation}
where $F_2^{as}(x,Q^2)=F_2(x,Q^2\gg |B(x)|)$ and the remaining
$Q^2$-dependence in $F_2^{as}(x,Q^2)$ is to be attributed to QCD logarithmic 
corrections only. 

Power corrections can be incorporated in the first term of Eq.~(\ref{a1}) by 
using a different scaling variable, 
\begin{equation}
\hat x =\phi (x,Q^2)=x+\frac{b(x)}{Q^2}+\cdots,
\label{a3}
\end{equation}  
so that
\begin{equation}
F_2(x,Q^2)=F_2\left (\phi^{-1}(\hat x,Q^2),Q^2\right )
\simeq F_2^{as}(\hat x,Q^2)\,. 
\label{a2}
\end{equation}
The coefficient $B$, which determines the value of the power 
correction in Eq.~(\ref{a1}), is thus related to the structure function by  
$B(x)=b(x)\, \partial F_2^{as}(x,Q^2)/\partial x$. In fact, an analysis of data 
in terms of an appropriate scaling variable appears to be more convenient, 
than the direct evaluation of power corrections. 

In general, the power corrections are generated by confining  
interaction of partons in the final state.
At first sight, such an interaction  should influence the 
structure function very drastically. Consider for instance the example of
two {\em nonrelativistic} ``quarks" of mass $m$ interacting via a harmonic
oscillator potential\cite{greenb}. These quarks are never free
and therefore the system in the final state possesses a discrete spectrum.
As a result the structure function, $F(q,\nu )$, as a function of 
the energy transfer $\nu$, is
given by a sum of $\delta$ - functions.
Obviously, it looks very different from the structure
function obtained in the impulse approximation, 
which considers the struck parton as 
a free particle in the final state. This paradox can be resolved by 
introducing a (nonrelativistic) scaling
variable $y$ 
\begin{equation}
y=-\frac{|{\mbox{\boldmath $q$}}|}{2}+\frac{m\nu}{|{\mbox{\boldmath $q$}}|}, 
\label{c2}
\end{equation}
Then expanding the structure function  ${\cal F}(q,y )\equiv F(q,\nu )$ 
in powers of $1/q$, one finds in the limit $q\to\infty$ and $y=$const that the 
$\delta$-peaks merge to a smooth curve,  
${\cal F}(q,y )\to {\cal F}_0(y )$, which coincides with
a free parton response\cite{gr}. 
Although this result appears to confirm the
parton model picture, it does not imply that the interaction in the final
state is not important. The latter has been merely incorporated in
${\cal F}_0(y )$ by an appropriate choice of the scaling
variable $y$, as shown by Eqs.~(\ref{a1}-\ref{a3}). The remaining 
contribution from higher-order ($\sim 1/q$) terms are thus minimized. 
However, a non-optimal choice of the scaling variable could 
result in very large or even singular corrections  
to the structure function. One can anticipate that an
appropriate choice of of the scaling variable is especially relevant at 
large $x$, where lower-lying excitations should play an important role. 

A general analysis performed in the framework of Bethe-Salpeter equation 
shows that in an analogy with the nonrelativistic case\cite{greenb,gr} 
the higher twist terms from a local confining final state interaction 
and target mass effects can  be effectively 
accounted in the modified the struck quark propagator, where 
the quark mass has the same off-shell value before 
and after the virtual photon absorption\cite{gur}, 
\begin{figure}
\vspace{0.2cm}
\hspace{0.4cm}
\epsfxsize=15cm
\epsfysize=15cm
\epsffile{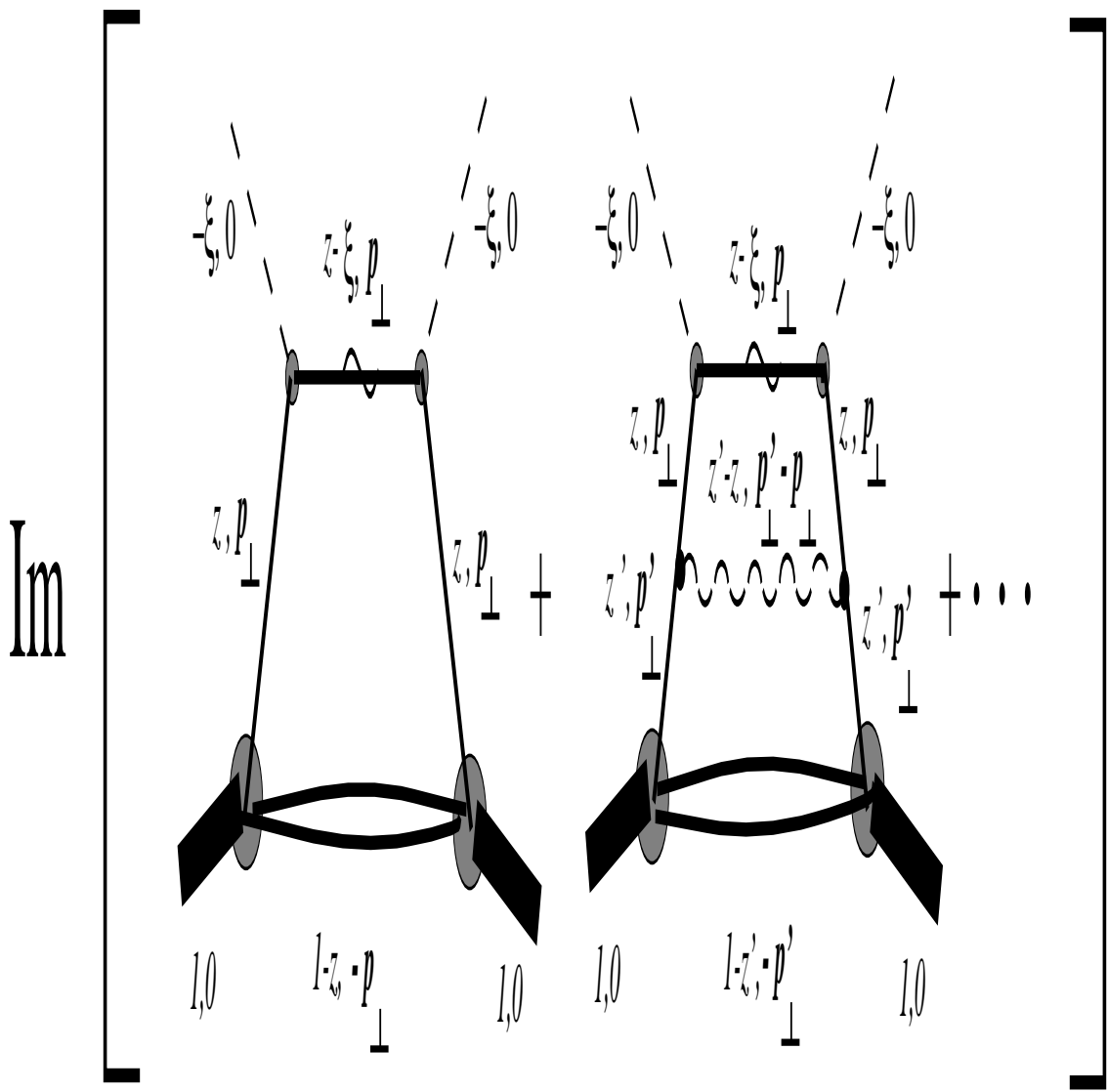}
\caption {Diagrammatic representation of the leading 
contributions to the structure function using light-cone variables. 
Quarks and gluons are shown by solid and wavy lines respectively.
The modified propagators are marked by ``$\sim$".} 
\label{fig1}
\end{figure}

As a result, the Bjorken 
scaling variable $x$ is replaced by a new scaling variable 
$\bar x\equiv\bar x(x,Q^2)$, which is   
the light-cone fraction of the {\em off-shell} struck quark.
Explicitly,
\begin{equation} 
\bar x=\frac{x+\sqrt{1+4M^2x^2/Q^2}-
\sqrt{(1-x)^2+4m_s^2x^2/Q^2}}
{1+\sqrt{1+ 4M^2x^2/Q^2}},
\label{a5}
\end{equation} 
where $M$ is the target mass and $m_s$ is the invariant mass of 
spectator partons (quarks and gluons). For $Q^2\to \infty$ or for $x\to 0$
the variable $\bar x$ coincides with the Bjorken variable $x$.
However, at finite $Q^2$ these variables are quite different. 

It follows from Eq. (\ref{a5}) that $\bar x$ depends on the invariant
spectator mass, $m_s$. In terms of light-cone variables,  
Fig. 1, it can be written as
\begin{equation}
m_s^2=m_0^2+(1-z)\frac{
({\mbox{\boldmath $p'$}}_{\perp}-{\mbox{\boldmath $p$}}_{\perp})^2}
{z'-z}+(z'-z)\frac{m_0^2+{\mbox{\boldmath $p'$}}^2}{1-z'}+
{\mbox{\boldmath $p'$}}_{\perp}^2-{\mbox{\boldmath $p$}}_{\perp}^2,
\label{c3}
\end{equation}
where $m_0$ is the diquark mass, and $1\geq z'\geq z$. We approximate 
$m_s$ as an effective spectator mass 
depends only on external momenta. Since $z\to\bar x\simeq x$ and 
$z'\sim z$, one gets from Eq. (\ref{c3})
\begin{equation} 
m_s^2\simeq m_0^2+C(x,Q^2)(1-x)
\label{a6}
\end{equation}

Eq. (\ref{a6}) for the invariant spectator mass looks quite appealing 
apart from its relation to Eq. (\ref{c3}). Indeed, $x=1$ corresponds to 
elastic scattering, when no gluons are emitted. Therefore in this
case the spectator is represented by a diquark. When 
$x$ decreases, gluons are emitted and $m^2_s$ increases $\propto (1-x)$.    
The coefficient $C(x,Q^2)$ in Eq. (\ref{a6})  
determines the rate of increase of the spectator mass with $Q^2$ and $x$. 
It can be found self-consistently from the evolution 
equation. However, when $x\sim 1$, one can take  
$C(x,Q^2)\simeq C(1,Q^2)\simeq$ const, because of  
$Q^2$-dependence of the spectator mass is less important than its 
$x$-dependence near the elastic threshold. 
We roughly estimated the value of $C$ by using the Weizs\"acker-Williams 
or ``equivalent photon" approximation, utilized in Ref.\cite{jaf}  
for derivation of the evolution equation. One finds from\cite{jaf} that 
the light-cone fraction of the ``equivalent" gluon, 
$z-z'$, (Fig. 1) is of order $\alpha_s \ln (Q^2/Q_0^2)$
in the region of large $x$. 
However, the probability of the gluon emission   
is also about the same order of magnitude. Then, as follows from 
Eq. (\ref{c3}), $C\sim\langle 
({\mbox{\boldmath $p'$}}_{\perp}-{\mbox{\boldmath $p$}}_{\perp})^2
\rangle$, so that one could expect to find $C$ on the scale of (GeV)$^2$.
In the following we regard  
it as a phenomenological parameter, determined from the data.

Let us consider the nucleon structure functions in the region of large $x$, 
where the power corrections to the scaling are dominant. 
At present, the only available large-$x$ data for proton and 
deuteron structure functions, $F_2^p(x,Q^2), F_2^d(x,Q^2)$, 
are the SLAC data\cite{slac1,slac2,slac3,slac4},
taken at moderate values of momentum transfer, $Q^2 < 30$ (GeV/c)$^2$.
(The nucleon structure functions for higher values of momentum
transfer ($Q^2\leq 230$ (GeV/c)$^2$) are extracted 
from BCMDS\cite{bcd1} and NMC\cite{nmc1} data, yet only for $x\leq 0.75$). 
The SLAC data for the proton and deuteron structure functions  
for $x\geq 0.7$ and $5 < Q^2 < 30$ (GeV/c)$^2$ are shown in 
Fig. 2 as a function of $x$. Also shown is the value of $F_2^p(x,Q^2)$ and 
$F_2^d(x,Q^2)$ for $Q^2$= 230 (GeV/c)$^2$ and $x=0.75$ taken from
the BCDMS data\cite{bcd1}. The data points close 
to the region of resonances were excluded by a requirement that the 
invariant mass of the final state  
$(M+\nu )^2-{\mbox{\boldmath $q$}}^2$ is greater than 
$(M+\Delta)^2$, where $\Delta=$ 300 MeV. In addition, we excluded the data  
points with $x>0.9$ from the deuteron structure only (Fig. 2b).     
The reason is that the deuteron structure function can not be 
represented as an average of the proton and neutron structure functions for 
$x> 0.9$. Indeed, the calculations of Melnitchouk {\em et al.}\cite{mel}
show that the ratio  $2F_2^d/(F_2^p+F_2^n)$ is about 1.13 for 
$x=0.9$ and $Q^2$=5 (GeV/c)$^2$, and it rapidly increases for $x>0.9$. 
However, for $x<0.85$, this ratio is within 5\% of unity\cite{not1}.

One finds from Fig.2 that the structure functions show 
no scaling in the Bjorken variable $x$. Also, very poor scaling is  
obtained when the data are plotted as a function of the Nachtmann 
variable $\xi$\cite{slac4}. However, the situation is 
different if we display the same data as a
function of the variable $\bar x$, Eq.~(\ref{a5}). It appears 
that the scaling in the $\bar x$-variable is strongly dependent 
on the value of diquark mass, $m_0$, Eq.~(\ref{a6}), but is much less 
sensitive to variation of the coefficient $C$.
For instance, the data as a function of $\bar x$  
display very poor scaling for $m_0 > 600$ MeV, i.e. 
by considering the spectator as build up from constituent quarks.
\begin{figure}
\vspace{0.2cm}
\hspace{0.4cm}
\epsfxsize=10cm
\epsfysize=20cm
\epsffile{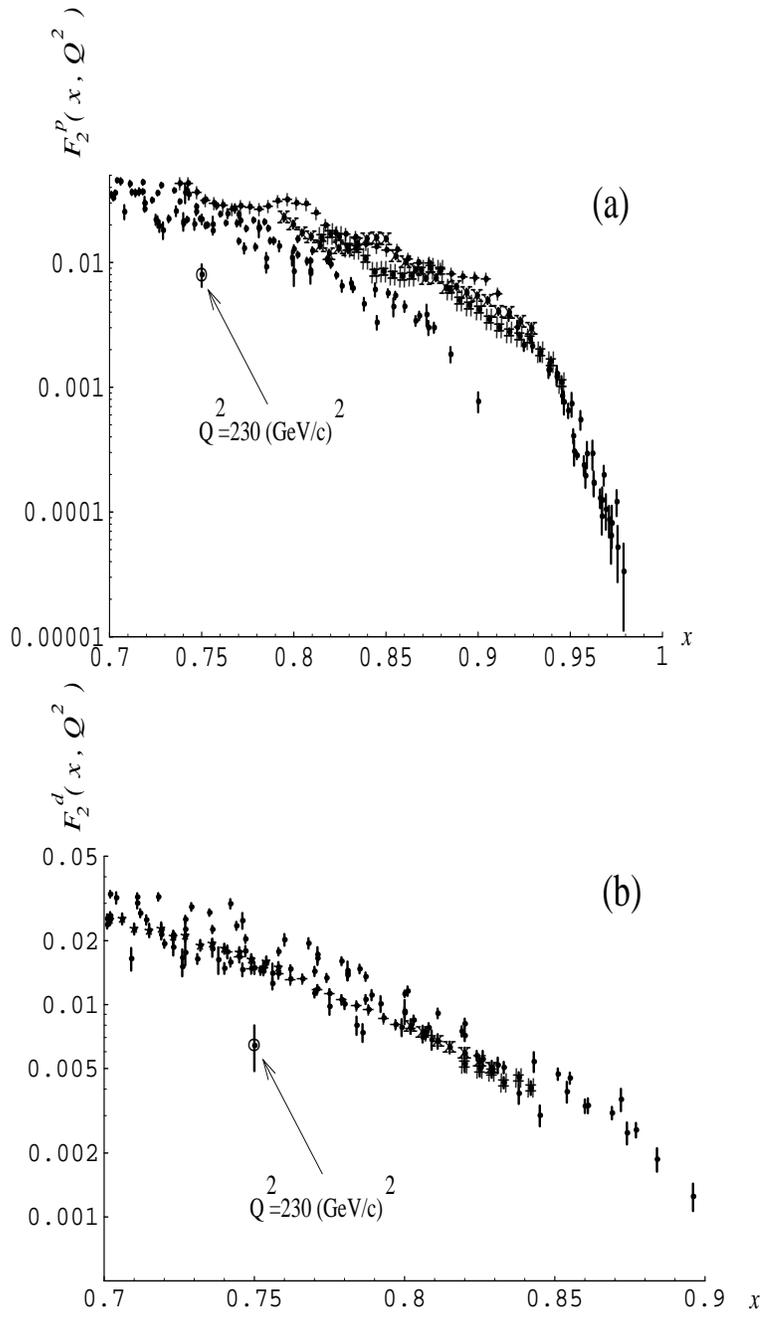}
\caption{The SLAC data[5,6,7,8]
($5\leq Q^2\leq 30$ (GeV/c)$^2$)
for proton (a) and deuteron (b), 
are shown as a function of the Bjorken variable
$x$. Three high-statistics data sets[7]  
for $Q^2\simeq$5.7, 7.6, and 9.5 (GeV/c)$^2$ are marked by  
``+", ``x", and ``{\#}" respectively.
The point at $Q^2=230$ (GeV/c)$^2$ and $x=0.75$ is from ref.[9].}
\label{fig2}
\end{figure}

\begin{figure}
\vspace{0.2cm}
\hspace{0.4cm}
\epsfxsize=10cm
\epsfysize=20cm
\epsffile{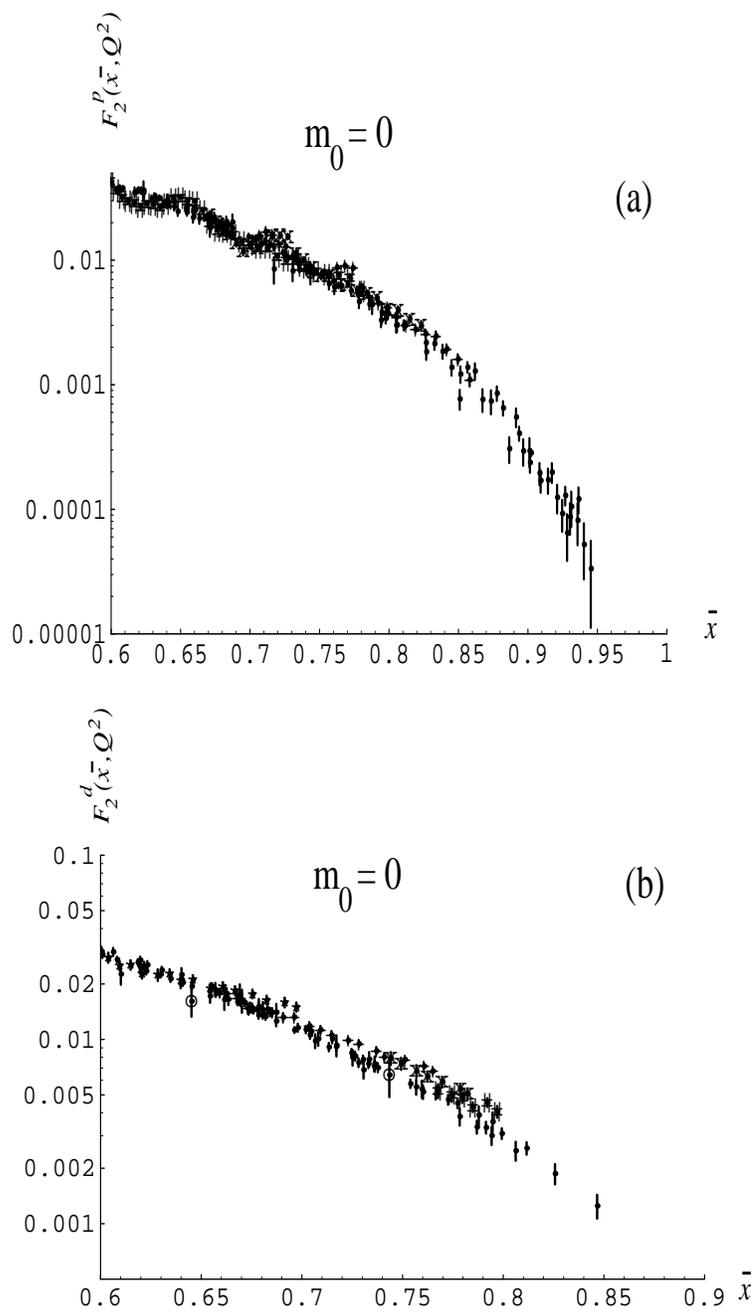}
\caption{The data of Fig. 2 are shown as function of the 
$\bar x(x,Q^2)$---the scaling variable of Eq.~(5) --- 
assuming  $m_0=0$ MeV and $C$ = 3(GeV)$^2$ for the spectator mass $m_s$,
Eq.~(7).}
\label{fig3}
\end{figure} 
On the other hand, the scaling is very good both
for the proton and deuteron data, by taking
$m_0=0$, i.e. by considering the spectator build up by current
quarks\cite{gur}. The results are shown in Fig. 3, where the data are
plotted as a function of $\bar x$ for $m_0=0$ and $C$ = 3 (GeV)$^2$. 
Note, that the high-$Q^2$ data points from BCDMS data\cite{bcd1} are
very close to the SLAC data points, taken at much lower values of $Q^2$.

Now by using the scaling variable $\bar x$, Eq.~(\ref{a5}) for $m_0=0$  
we can analyze the nucleon structure functions for smaller values 
of $x$, where both power and logarithmic corrections 
to the Bjorken scaling play an important role. In this region  
($0.35 < x < 0.75$) the existing BCDMS\cite{bcd1} and NMC\cite{nmc1} data 
are extended up to much larger values of momentum transfer  
than the previously considered high-$x$  
SLAC data. It allows us to check our predictions in a wide $Q^2$ range.
The results\cite{gmt} are shown in Figs. 4 and 5 for proton and deuteron 
structure functions respectively. The data points are from  SLAC and BCDMS data 
bins\cite{slac1,bcd1}. The dotted lines show the $Q^2$-dependence of the 
structure functions due to power corrections only.
\begin{figure}
\vspace{0.2cm}
\hspace{0.4cm}
\epsfxsize=15cm
\epsfysize=25cm
\epsffile{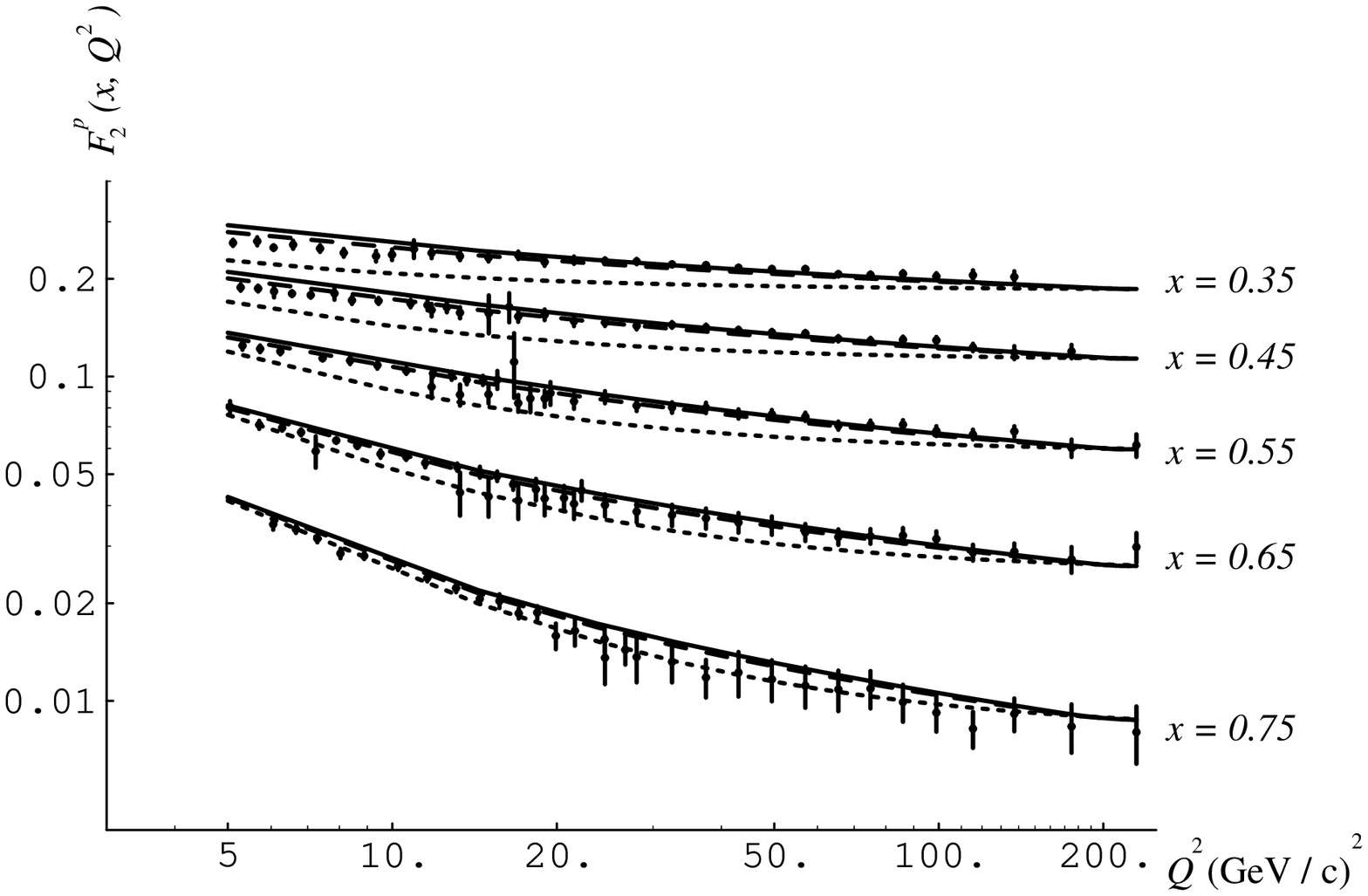}
\caption{The proton structure function $F_2^p(x,Q^2)$ is shown
as a function of $Q^2$ at different $x$-values. The dotted lines include power
corrections only. They are evaluated according to and the
scaling variable $\bar x$ of Eqs.~(5), (7) with $m_0=0$
and $C$ = 3 (GeV)$^2$. The additional QCD logarithmic corrections evaluated at
NLO for different $\Lambda$ scales are shown by the dashed 
($\Lambda = 100$ MeV) and continuous lines ($\Lambda = 200$ MeV).}
\label{fig4}
\end{figure} 
The total $Q^2$-dependence of structure functions due to the power and
the logarithmic NLO corrections, is shown by the dashed 
and continuous lines for $\Lambda$ = 100 MeV and $\Lambda$ = 200 MeV
respectively. The QCD (logarithmic) evolution corrections are taken 
into account at Next-to-Leading Order (NLO)
evolving {\em back}, in $Q^2$, the structure functions
starting from an asymptotic value of momentum transfer where the condition
$F_2(x,Q^2) \simeq F_2^{as}(x,Q^2)$ (cf. Eq.~(\ref{a1})) is 
fulfilled (in the present case we choose 
$Q^2 = 230$ (GeV/c)$^2$, which is the highest value 
of the momentum transfer in the BCDMS data\cite{bcd1}).
\begin{figure}
\vspace{0.2cm}
\hspace{0.4cm}
\epsfxsize=15cm
\epsfysize=25cm
\epsffile{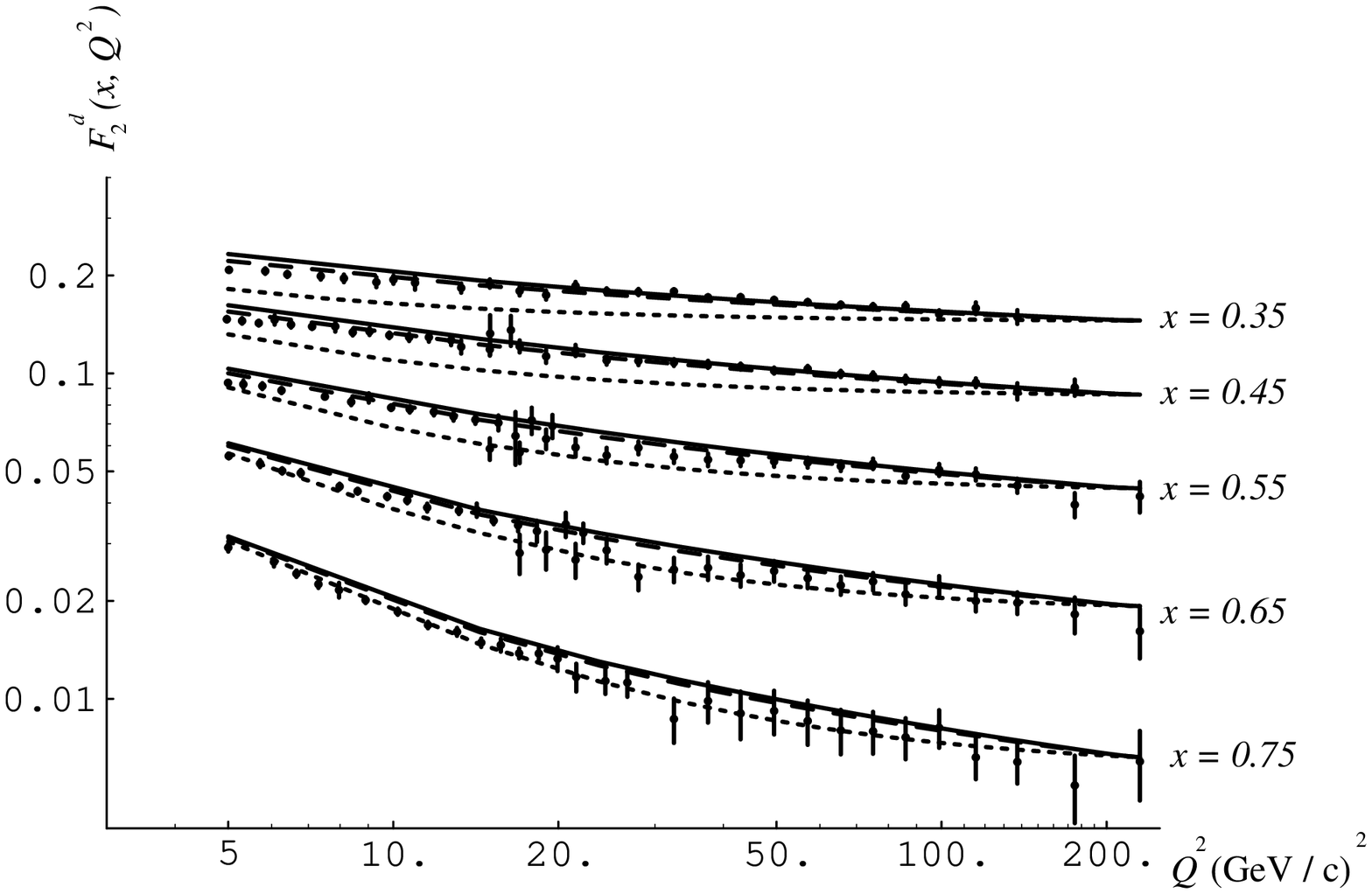}
\caption{As in Fig. 4 for the deuteron structure function
$F_2^d(x,Q^2)$.}
\label{fig5}
\end{figure} 

One finds from Figs. 4 and 5 that the experimental data are reproduced 
in a large $Q^2$-range for both values of $\Lambda$, 
although the agreement is slightly better for $\Lambda$ = 100 MeV.  
In addition, since the results are strongly dependent on the spectator mass 
({\ref{a6}), it is remarkable that the same parameters $m_0=0$ and 
$C$ = 3 (GeV)$^2$ do reproduce the $Q^2$-behavior of the structure functions 
both for large and moderate $x$-values.

The above analysis of the nucleon structure functions allows us to extract 
from data the previously unattainable information on the asymtotic 
ratio of $F_2^n(x)/F_2^p(x)$ at large $x$, directly related to $u$ and 
$d$-quarks distribution. Indeed, according to the quark-parton model, 
the structure functions of hadrons 
in the Bjorken limit ($Q^2={\mbox{\boldmath $q$}}^2 -\nu^2\to \infty$ and 
$x=Q^2/2M\nu$=const) are directly related to the parton distributions 
q$_i(x)$. For instance  
\begin{equation}
F_2(x,Q^2)\to F_2(x)=\sum_i e_i^2x{\rm q}_i(x),
\label{b1}
\end{equation}
where the sum is over the partons, whose charges are $e_i$. In the region 
$x\to 1$, the contribution of sea quarks can be neglected. Then assuming the 
same distribution of the valence quarks, one easily finds from 
Eq. (\ref{b1}) that the neutron-to-proton ratio 
$F_2^n(x)/F_2^p(x)$ approaches 2/3 for $x\to 1$. If, however, the quark 
distributions are different, 
one can establish only upper and lower limits for this ratio,
$1/4<F_2^n/F_2^p<4$, which follow from isospin invariance\cite{nacht1}. 

In order to check the parton model predictions
one needs to take the structure functions at high $Q^2$, 
since the higher-twist corrections to the scaling are very important   
at high-$x$ region. At present, high $Q^2$ structure 
functions ($Q^2\simeq 250$ (GeV/c)$^2$) extracted from 
BCDMS\cite{bcd1} and NMC\cite{nmc1} data are available only 
for $x\leq 0.7$. The ratio  $F_2^n(x)/F_2^p(x)$  
obtained from an analysis of these data\cite{bcd2} shows steady 
decrease with $x$. Thus, it is usually assumed that this ratio would reach 
its lower bound, $F_2^n/F_2^p\to 1/4$, for $x\to 1$ 
(although the recent analysis\cite{mel1} suggests that this ratio 
would be larger than  1/4 for $x\to 1$). This
corresponds to $d(x)/u(x)\to 0$ for $x\to 1$, where $d(x)$ and $u(x)$ 
are the distribution functions for up and down quarks in the proton.

Since our scaling variable $\bar x$ takes effectively into account the 
higher-twist corrections, it allows us to reach the asymptotic limit 
already at moderate values of momentum 
transfer, $Q^2 < 30$ (GeV/c)$^2$. 
Thus, $F_2^p$ and $F_2^d$ extracted from the SLAC
data at large values of $x$\cite{slac1,slac2,slac3,slac4}, Fig. 3,  
would represent the asymptotic structure functions: $F_2(\bar x)=
F_2(x)\equiv F_2(x,Q^2\to\infty)$, which can be used as an input 
to determine the ratio $F_2^n(x)/F_2^p(x)$.  
For this purpose we parametrize 
the asymptotic structure functions as 
$F_2^{p,d}(x)=\exp (-\sum_{i=0}^4 a_ix^i)$ 
and determine the parameters $a_i$ from the best fit to the data in 
Figs. 3a, 3b. The resulting $F_2^n/F_2^p$ ratio is shown 
in Fig. 6 by the solid line. The dotted lines are the error bars
on the fit, which combine 
statistical and systematic uncertainties.  
The dashed line corresponds to $F_2^n/F_2^p=2/3$.
For a comparison, we show by the dot-dashed line 
a polynomial extrapolation of this ratio to large $x$, 
obtained from BCDMS and NMC data by assuming that $F_2^n/F_2^p\to 1/4$
for $x\to 1$\cite{bcd2}. 
structure functions
\begin{figure}
\vspace{0.2cm}
\hspace{0.4cm}
\epsfxsize=15cm
\epsfysize=25cm
\epsffile{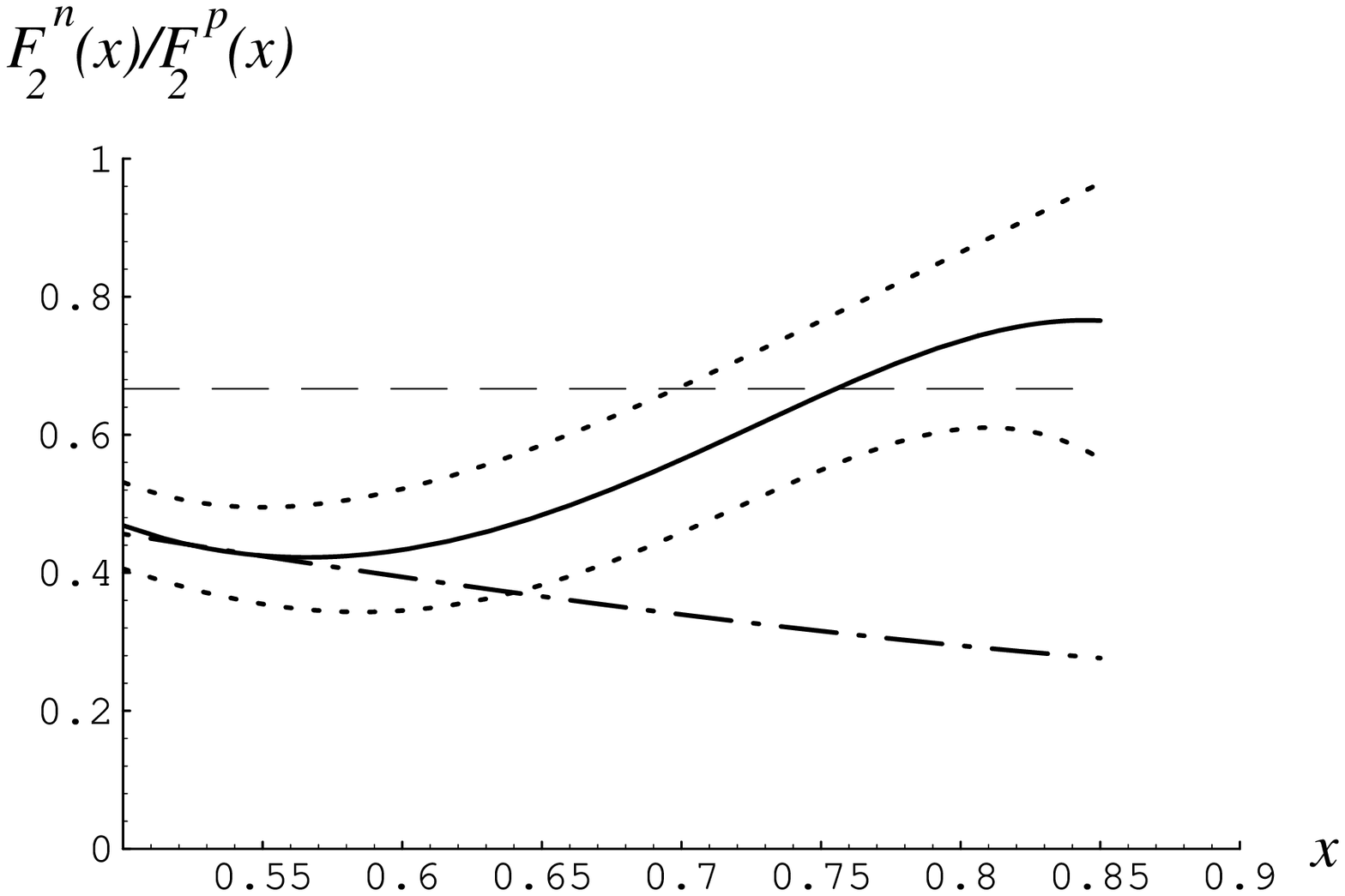}
\caption{Neutron-to-proton structure function ratio at large $x$.} 
\label{fig6}
\end{figure} 
Our results shown in Fig. 6 demonstrate that contrary to earlier 
expectations, the ratio
$F_2^n/F_2^p$ does not approach its lower bound, but increases up to 
approximately 2/3. The latter is the 
quark model prediction, assuming the same distributions for each 
of the valence quarks. 
The accuracy of our results will be checked in future experiments, 
which will provide high $Q^2$ data for the structure functions at 
large $x$.

The described above method for evaluation of the higher-twist 
corrections using modified scaling variable can be very 
useful for not only for an analysis of high-$x$ nucleon structure 
functions, but also for a treatment of many different problems, 
where the interaction of partons in the final state becomes important.
In particular, I would like to mention the sum rules, where the 
role of higher-twist effects remains an open problem, and 
the structure functions at $x\to 0$. In this region the 
spectator mass $m_s$ is not described by Eq. (\ref{a6}), valid  
for large $x$. If $m_s^2\propto 1/Q^2$ for $x\to 0$, 
then $(\bar x -x)/x\propto 1/Q^2$, Eq.~(\ref{a5}), and the role of higher-twist 
correction would be substantial in this region too.


\begin{thebibliography}{99}

\bibitem{greenb}  O.W. Greenberg, Phys. Rev. D{\bf 47}, 331 (1993);
\bibitem{gr}S.A. Gurvitz and A.S. Rinat, Phys. Rev. C{\bf 47}, 2901 (1993).

\bibitem{gur}  S.A. Gurvitz, Phys. Rev. D{\bf 52}, 1433 (1995).

\bibitem{jaf}  R.L. Jaffe, in {\em Los Alamos School on Relativistic 
Dynamics and Quark-Nuclear Physics}, ed. M.B. Jackson and A. Picklesimer.
(John Wiley and Sons, New York, 1985).

\bibitem{slac1} L.W. Whitlow, Ph.D. Thesis, Stanford University, 1990,          
SLAC-REPORT-357 (1990). 

\bibitem{slac2}  L.W. Whitlow {\em et al.},  Phys. Lett. B{\bf 282}, 
475 (1992).

\bibitem{slac3} S.E. Rock {\em et al.}, Phys. Rev. D{\bf 46}, 24 (1992).

\bibitem{slac4} P.E. Bosted {\em et al.}, Phys. Rev. D{\bf 49}, 3091 (1994).

\bibitem{bcd1} BCDMS Collab., A.C. Benvenuti {\em et al.}, 
Phys. Lett. B{\bf 223}, 485 (1989); Phys. Lett. B{\bf 237}, 592 (1989).

\bibitem{nmc1} NMC Collab., P. Amaudruz {\em et al.}, Phys. Lett. B{\bf 295}, 
159 (1992).

\bibitem{mel}  W. Melnitchouk, A.W. Schreiber and A.W. Thomas, 
Phys. Lett. B{\bf 335}, 11 (1994).

\bibitem{not1} Similar small binding and Fermi motion effects 
in the deuteron structure function were also found in a recent  
phenomenological analysis of J. Gomez {\em et al.}, 
Phys. Rev. D{\bf 49}, 4348 (1994). 

\bibitem{gmt} S.A. Gurvitz, A. Mair and M. Traini, Phys. Lett. B, in press. 

\bibitem{nacht1} O. Nachtmann, Nucl. Phys. B{\bf 38}, 397 (1972).

\bibitem{bcd2}  BCDMS Collab., A.C. Benvenuti {\em et al.},  
Phys. Lett. B{\bf 237}, 599 (1989).

\bibitem{mel1} W. Melnitchouk and A.W. Thomas, Phys. Lett. B{\bf 377}, 11 (1996).

\end{thebibliography}
\end{document}